\documentclass[conference]{IEEEtran}
\IEEEoverridecommandlockouts
\usepackage{cite}
\usepackage{amsmath,amssymb,amsfonts}
\usepackage{algorithmic}
\usepackage{graphicx}
\usepackage{textcomp}
\usepackage{xcolor}
\usepackage{algorithm}
\usepackage{algorithmic}
\usepackage{makecell}
\usepackage{multirow}
\usepackage{subfig}
\usepackage[hyphens]{url}
\usepackage[colorlinks,urlcolor=blue,linkcolor=blue,citecolor=blue]{hyperref}
\def\BibTeX{{\rm B\kern-.05em{\sc i\kern-.025em b}\kern-.08em
    T\kern-.1667em\lower.7ex\hbox{E}\kern-.125emX}}
\begin{document}

\title{Demand Modeling for Advanced Air Mobility\\
\thanks{This material is based upon work supported by the NASA Aeronautics Research Mission Directorate (ARMD) University Leadership Initiative (ULI) under cooperative agreement number 80NSSC23M0059.}
}

\author{\IEEEauthorblockN{Kamal Acharya}
\IEEEauthorblockA{\textit{Department of Information Systems} \\
\textit{University of Maryland Baltimore County}\\
Baltimore, MD, US \\
kamala2@umbc.edu}
\and
\IEEEauthorblockN{Mehul Lad}
\IEEEauthorblockA{\textit{Department of Information Systems} \\
\textit{University of Maryland Baltimore County}\\
Baltimore, MD, US \\
du72811@umbc.edu}
\and
\IEEEauthorblockN{Liang Sun}
\IEEEauthorblockA{\textit{Department of Mechanical Engineering} \\
\textit{Baylor University}\\
Waco,TX, US \\
liang\_sun@baylor.edu}
\and
\IEEEauthorblockN{Houbing Song}
\IEEEauthorblockA{\textit{Department of Information Systems} \\
\textit{University of Maryland Baltimore County}\\
Baltimore, MD, US \\
songh@umbc.edu}
}
\maketitle

\begin{abstract}
In recent years, the rapid pace of urbanization has posed profound challenges globally, exacerbating environmental concerns and escalating traffic congestion in metropolitan areas. To mitigate these issues, Advanced Air Mobility (AAM) has emerged as a promising transportation alternative. However, the effective implementation of AAM requires robust demand modeling. This study delves into the demand dynamics of AAM by analyzing employment based trip data across Tennessee's census tracts, employing statistical techniques and machine learning models to enhance accuracy in demand forecasting. Drawing on datasets from the Bureau of Transportation Statistics (BTS), the Internal Revenue Service (IRS), the Federal Aviation Administration (FAA), and additional sources, we perform cost, time, and risk assessments to compute the Generalized Cost of Trip (GCT). Our findings indicate that trips are more likely to be viable for AAM if air transportation accounts for over 70\% of the GCT and the journey spans more than 250 miles. The study not only refines the understanding of AAM demand but also guides strategic planning and policy formulation for sustainable urban mobility solutions. The data and code can be accessed on GitHub. \footnote{Github: \url{https://github.com/lotussavy/IEEEBigData-2024.git }}

\end{abstract}


\section{Introduction}
The exponential growth of urban populations and the geographic dispersal of employment hubs have escalated demand for efficient transportation solutions, impacting both workforce mobility and economic productivity. Traditional ground-based transit systems, despite their established infrastructure, struggle to meet the increasing demand, leading to extended travel times, higher fuel consumption, and substantial economic losses from congestion. According to the 2023 Urban Mobility Report by the Texas A\&M Transportation Institute\footnote{2023 URBAN MOBILITY REPORT: \url{https://static.tti.tamu.edu/tti.tamu.edu/documents/mobility-report-2023.pdf }. Last accessed on [May 9\textsuperscript{th} 2024]},"The total travel delay increased from 4.5 billion hours in 2020 to 8.5 billion hours in 2022. Similarly, wasted fuel increased from 1.9 billion gallons to 3.3 billion gallons. The congestion costs increased significantly, from \$113 billion in 2020 to \$224 billion in 2022 and the travel volume increased from 1.38 trillion miles traveled in 2020 to 1.515 trillion miles in 2022." The solution to all the problems is Advanced Air Mobility (AAM) which is an emerging form of aviation that includes electric, autonomous aircraft for transportation.

AAM being advanced form of aviation is used for transportation in several ranges and on the basis of which it is divided into two different categories Urban Air Mobility (UAM) and Regional Air Mobility (RAM). At their foundation, UAM and RAM share numerous similarities, such as the recognition of the benefits of advanced electrified propulsion architectures, increased utilization, and Single Vehicle Operation (SVO)/autonomy in drastically reducing operational costs. However, UAM and RAM diverge significantly in the markets they serve and their infrastructural needs. UAM primarily uses electric vertical take-off and landing (eVTOL) aircraft to access urban cores through newly developed vertiports, whereas RAM utilizes the extensive network of underutilized existing airports across the United States, eliminating the necessity for vertical takeoff and landing capabilities\cite{antcliff2021regional}. Operation range of UAM is less than 150 km whereas for RAM this range is 150-800km\cite{mckinseyShorthaulFlying}. With rapid advancements in electric propulsion and autonomous systems, the AAM market emphasizes sustainability and efficiency to address urban congestion and environmental issues. Implementing AAM services presents several challenges and interdisciplinary constraints that companies must address, including decisions regarding the placement of air taxi stations, effective routing and coordination of thousands of air taxis across the network, and advanced data analytics to forecast demand in real-time. Companies must work to minimize passenger commute time and costs while maximizing the efficiency of on-demand ride-sharing. Operational issues such as developing pricing strategies, evaluating first and last-mile delivery options, and monitoring critical metrics like air taxi battery status and maintenance needs also need to be addressed.

This study specifically focuses on modeling the demand for AAM as a passenger transportation mode. To achieve this, the initial step involves assessing the demand for AAM services by analyzing existing trip demand data and identifying trips that qualify for AAM. The trip demand is examined at the census tracts level within Tennessee. Utilizing various datasets from sources such as the Bureau of Transportation Statistics (BTS), Internal Revenue Service (IRS),LEHD Origin-Destination Employment Statistics (LODES) dataset and the Federal Aviation Administration (FAA), we conduct cost, time, and risk modeling. Based on these models, we calculate the Generalized Cost of Trip (GCT) to predict whether a trip will be served by ground transportation or AAM. We have considered all the 70 airports in the Tennessee as hubs for AAM operations.

This paper contributes to the growing body of research on AAM by addressing several key challenges unique to this sector. 
\begin{itemize}
    \item First, it highlights the specific demand dynamics that differentiate AAM from both traditional aviation and public transportation, filling a gap in the literature that has often focused on urban air mobility (UAM) within singular city contexts. 
    \item Second, by introducing a novel cost and time modeling framework that factors in both regional and urban contexts, this study demonstrates how AAM can serve as a complementary transportation mode, optimized for varying trip distances and user needs. 
    \item Finally, we present a probabilistic demand model that incorporates uncertainties in traveler preferences, offering insights into how AAM might adapt to both routine and fluctuating demand patterns.
\end{itemize}

The remainders of this paper are organized in the following manner: A review of prior research is presented in the related work, followed by the methodology. Subsequently, an evaluation and discussion of our findings are detailed, leading to the final conclusions.

\section{Related Work}
The study by Rajendran \cite{rajendran2019insights} introduces a two-phase clustering methodology for optimizing air taxi station placements in urban settings. The first phase, termed "warm start," innovatively generates initial clustering seeds by integrating multimodal transportation data, focusing on existing transportation hubs and patterns to predict potential high-demand areas for air taxis. The second phase involves an iterative clustering process that refines these initial placements by incorporating practical constraints such as estimated demand and operational feasibility. Potential passengers for air taxi services were identified using assumptions proposed in the Uber white paper\cite{holden2016fast} and additional assumptions by the researcher. The Davies-Bouldin Index (DBI)\cite{davies1979cluster} was employed as the validity metric for the clustering algorithm. The dataset used consists of 300 million records of yellow and green taxi trips spanning from January 2014 to December 2015.

In a subsequent study by Rajendran \cite{rajendran2021predicting}, the dataset from the 2019 study\cite{rajendran2019insights} was enhanced with environment-related features like temperature and wind speed. This study employs a data-driven approach to forecast the spatiotemporal demand for air taxi services in New York City using logistic regression, artificial neural networks, random forests, and gradient boosting. The methodology hinges on the aggregation and clustering of geographic and temporal data to refine the prediction model's accuracy, leveraging k-means clustering to segment the city into distinct areas for detailed demand analysis. This spatial segmentation facilitates the exploration of demand patterns across different city regions, enhancing the strategic allocation of air taxi resources. By categorizing demand into three levels—low, moderate, and high—the study aids in operational planning and resource distribution, thereby optimizing air taxi services to meet varying urban mobility needs effectively.

Ahmed\cite{ahmed2024demand} explores the potential for implementing UAM by forecasting market demand through deep learning methodologies. Utilizing a dataset of 150,000 ground taxi service records from the New York City Taxi and Limousine Commission from 2012 to 2020, the study employs deep learning models such as Long Short-Term Memory (LSTM), Gated Recurrent Unit (GRU), and Transformer to predict UAM demand. They introduced a parameter called the demand ratio, which is the product of passenger numbers and the distance, where a high ratio indicates high demand and vice versa. After calculating the demand ratio, various machine learning models were applied to predict it.

This paper\cite{tarafdar2019urban} investigates the feasibility of UAM landing sites and develops a fare model analysis for the Greater Northern California region, focusing on commuting passengers traveling to work from home and back. The study uses an all-electric, autonomous, multi-rotor aircraft with Vertical Takeoff and Landing (VTOL) capabilities. The researchers aim to optimize landing site locations using census tracts, estimate demand potential, and evaluate life-cycle costs. They analyze three scenarios with 200, 300, and 400 landing sites, concluding that 300 sites offer a practical balance between demand capture and economic viability. The study includes a fare model, proposing a base fare of \$20 for 20 miles plus \$1 per additional mile, targeting a competitive fare structure of \$1 to \$1.25 per passenger-mile. The analysis employs datasets like the National Household Travel Survey, American Community Survey, and Zillow data to refine site selection and cost estimates. Results suggest that UAM can significantly reduce travel time and congestion, but further research is needed to address public acceptance, reliability, and infrastructure development.

\begin{figure}[hbt!]
\centerline{\includegraphics[width=\columnwidth]{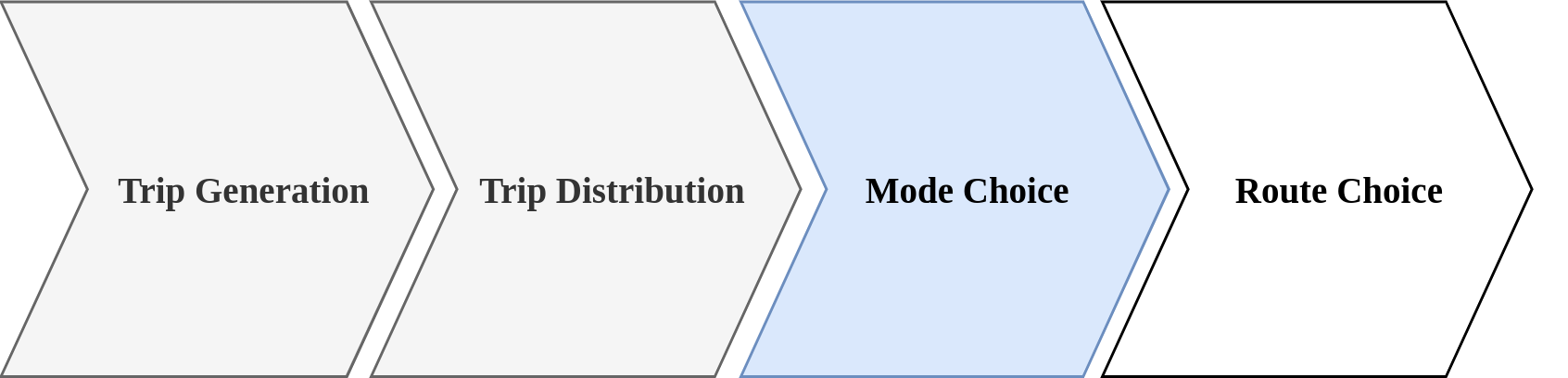}}
\caption{Four Step Model for travel demand modeling}
\label{fig:FSM}
\end{figure}

\renewcommand{\arraystretch}{1.15} 
\begin{table*}
\caption{\label{tab:Datasets} Datasets Used in the Research}
\centering
\begin{tabular}{|c|c|}
\hline
\textbf{Name }& \textbf{Details}\\ \hline

 BTS Monthly Traffic Dataset & Number of fatalities during Transportation \\ \hline

  USDA VSL Dataset & Monetary equivalent of reducing one death in population \\ \hline

LODES Dataset & OD pairs in employment basis \\ \hline

 BTS DB1BMarket Dataset & Ticket Price for the airlines\\ \hline

 IRS Standard Mileage Dataset & Standard Mileage Rates(cents/miles) \\ \hline
 
 FAA Airport Dataset& Block time of the flights\\
\hline
 BTS Inter-Airport Distance Dataset & Distance between the airports\\
\hline

 Open Source Routing Machine (OSRM) & Distance and Time of Ground transportation\\
 \hline

 US Bureau of Labor Statistics & Median hourly wages\\
\hline
\end{tabular}
\end{table*}

\begin{figure*}[hbt!]
\centering
\includegraphics[width=\textwidth]{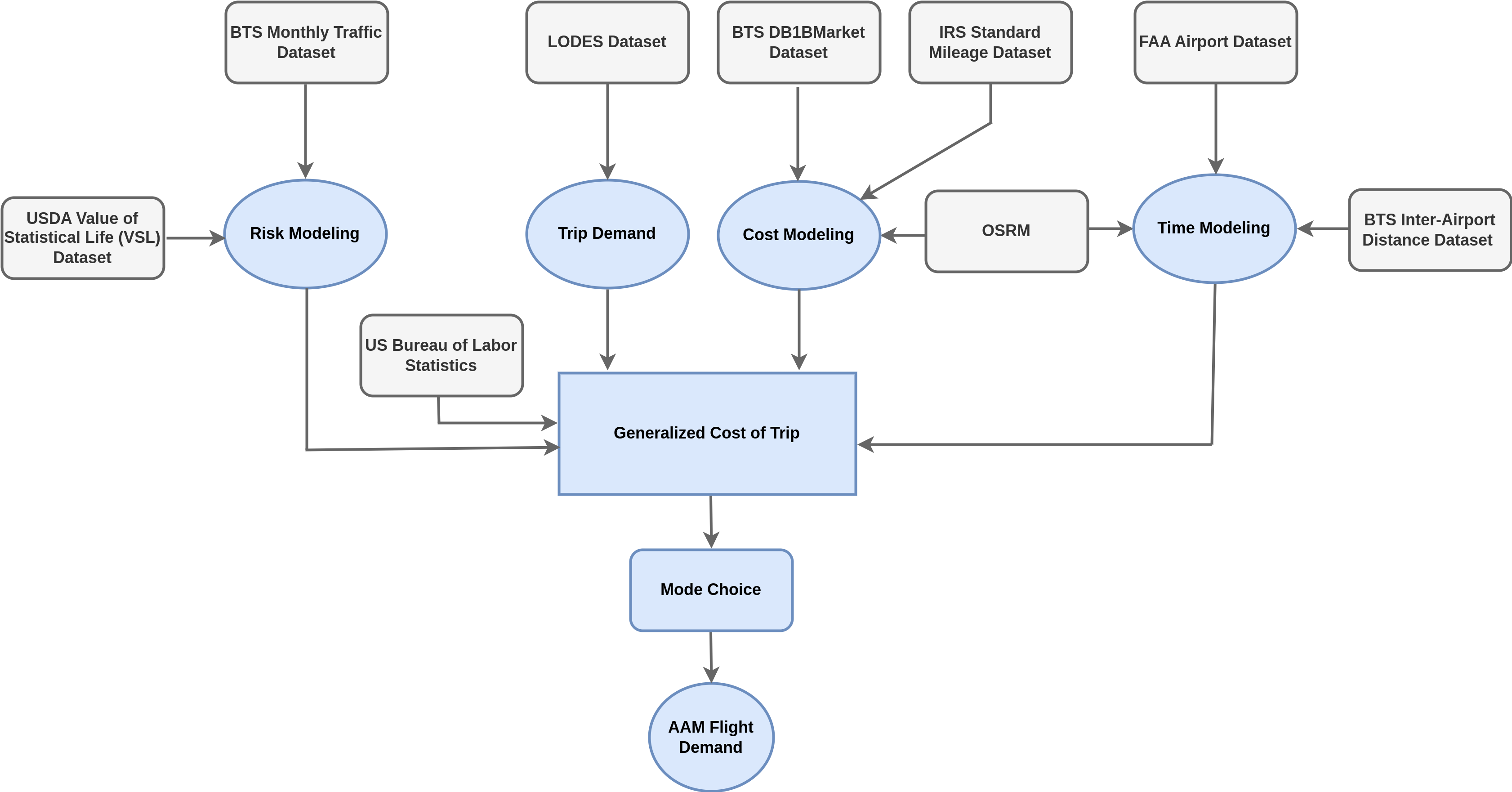}
\caption{Approach for AAM demand modeling}
\label{fig:Methodology}
\end{figure*}

\section{Methodology}
Flight demand forecasting of AAM  is an complicated approach to handle as it is affected by the number of different factors. Being new and flourishing technology, it is very complicated to analyse how many of the total trip request generated can be converted into the AAM demand. The whole process can be visualize through the Four Step Model(FSM)\cite{mcnally2007four} approach as shown in \autoref{fig:FSM}. The FSM in transportation planning encompasses four interconnected stages to predict traffic flows and travel demands. Trip generation estimates the number of trips originating and ending in different zones based on socio-economic and demographic factors, producing a count of trip productions and attractions. Trip distribution then allocates these trips between zones, using models like the gravity model to account for the influence of travel cost and distance, resulting in origin-destination matrices. Mode choice analyzes these matrices to determine the mode of transportation (car, bus, airlines etc.) travelers use, influenced by factors such as cost, time, and convenience. Final step, route choice assigns the trips to specific routes in the transportation network, optimizing for factors like travel time and congestion, particularly during different times of the day, to simulate actual traffic flow patterns.

Our focus is on the mode choice i.e. third step of FSM shown by light blue color in the \autoref{fig:FSM}. First two steps, trip generation and distribution are used as the supplements to help for the choosing the means of transportation. The last step of route choice represented by the white shade is out of the scope of this work as we focus only on the assignment of the generated trips to the mode of transportation.

Our approach for AAM demand modeling is shown in the \autoref{fig:Methodology}. It consists of several components for trip demand generation, risk modeling, cost modeling time modeling, GCT calculation and finally predicting the AAM demand. Details about these are provided in the upcoming sections. Various datasets that has been used in the research are mentioned in the \autoref{tab:Datasets}. 

The distances whenever used with the ground transportation are the driving distance we obtain from the OSRM\footnote{OSRM: \url{ https://project-osrm.org/}. Last accessed on [[July 29\textsuperscript{th} 2024]]} , which calculates the distance based on the actual road path, including turns and other deviations from a straight line. On the other hand distance used with the airlines is orthodromic or spherical distance which is the great circle distance between airports. To calculate the orthodromic or spherical distance between two points on the Earth's surface, we use the Haversine formula\cite{robusto1957cosine} given below:


\begin{equation}
\begin{split}
d = 2r \cdot \arcsin\Bigg( & \sqrt{\sin^2\left( \frac{\Delta \varphi}{2} \right) + \cos(\varphi_1) \cdot \cos(\varphi_2)} \\
& \cdot \sin^2\left( \frac{\Delta \lambda}{2} \right) \Bigg)
\end{split}
\end{equation}
where:
\begin{itemize}
    \item \( d \): Great-circle distance between the two points.
    \item \( r \): Radius of the Earth, \( r \) is approximately 6371 km (kilometers) or 3958.8 miles.
    \item \( \varphi_1 \) and \( \varphi_2 \): Latitudes of the two points in radians.
    \item \( \Delta \varphi = \varphi_2 - \varphi_1 \): Difference in latitude between the two points.
    \item \( \Delta \lambda = \lambda_2 - \lambda_1 \): Difference in longitude between the two points, where \( \lambda_1 \) and \( \lambda_2 \) are the longitudes of the points in radians.
\end{itemize}

\subsection{Region Selection}
Tennessee has been chosen as the focal point of our research due to its strategic position in advancing AAM initiatives. The state's innovation ecosystem is anchored by notable institutions like Whisper Aero and the University of Tennessee, which contribute significantly to research and technology development in advanced air mobility. On the geographical level we will be dealing with the trip demand on Census tract level. Census tracts are relatively permanent statistical subdivisions, ensuring consistency in data collection over time. However, adjustments are made to adapt to population shifts: tracts exceeding 8,000 inhabitants are split into smaller units with extended numeric codes, while those falling below 1,200 are merged with neighboring tracts and assigned new codes. Minor boundary corrections are permitted to reflect changes in local geography. These modifications are meticulously documented, allowing for accurate demographic and economic comparisons from decade to decade.\footnote{Census Tracts: \url{ https://www2.census.gov/geo/pdfs/education/CensusTracts.pdf }.Last accessed on [July 29\textsuperscript{th} 2024]}. We mapped the 1701 census tracts into one of the 70 airports to develop them as the hub for the AAM for the trip demand generated from each of these locations.

\begin{figure*}[hbt!]
    \centering
        \subfloat[Cost Regression Analysis For Airlines\label{fig:Cost}]{
        \includegraphics[width=0.98\columnwidth]{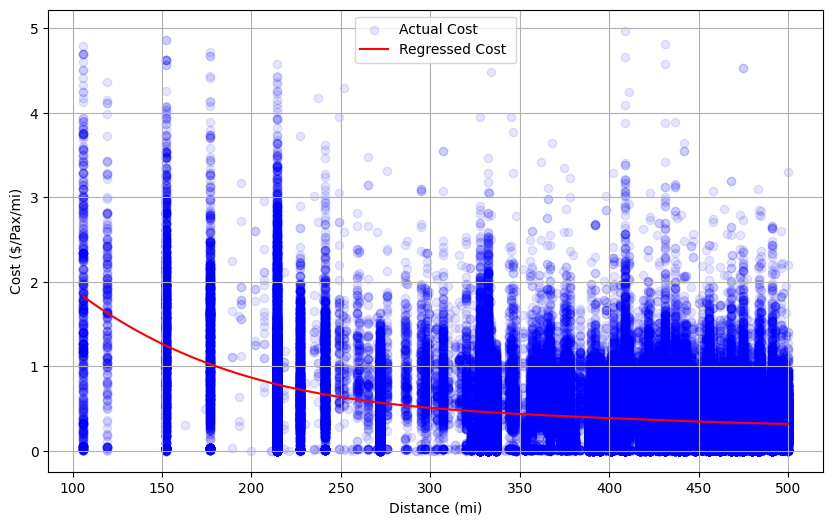}
    }
    \hfill
    \subfloat[Block Time Regression Analysis For Airlines\label{fig:BlockTime}]{
        \includegraphics[width=0.98\columnwidth]{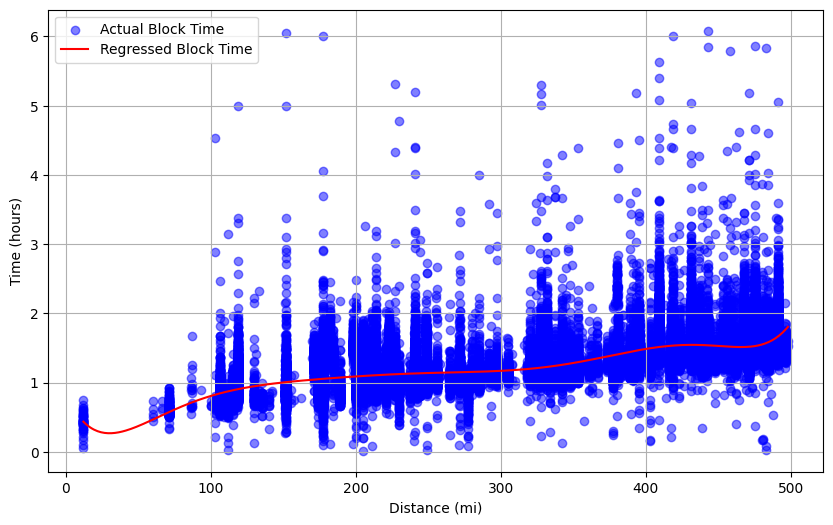}
    }
    \caption{Regression Analysis}
    \label{fig:reg}
\end{figure*}

\subsection{Trip Demand}
The first two steps of the FSM model are trip generation and trip distribution which serves as the supplements to our work for choosing the means of transportation.  The trip demand data from the Longitudinal Employer-Household Dynamics (LEHD) Origin-Destination Employment Statistics (LODES) dataset is specified at the census tract level, it is presumed that travel originates from the population centroid of the departure census tract and ends at that of the arrival census tract. We collect the Origin-Destination(OD) data for the year 2020. First we analyse all the trips data and later we filtered out the trip that satisfy the AAM requirement which is range of the distance should be from 150 to 800 KM for RAM and less than 150 KM for UAM. 

We have following assumptions for the trips:
\begin{enumerate}
    \item All trips generated from census tracts are generated from the centroid of population of census tracts.
    \item Ground transportation trip consists of distance travel from centroid of origin census tracts to centroid of destination census tracts.
    \item AAM trip consists of ground transportation from centroid of origin census tract to the nearest hub airport and then AAM flight to destination airports and finally ground transportation from destination airports to the centroid of destination census tract.
\end{enumerate}

\subsection{Cost Modeling}
 One critical factor is the cost estimation associated with each trip demand for each mode of transportation. Given that the trip demand data from the LODES dataset is specified at the census tract level, it is presumed that travel originates from the population centroid of the departure census tract and ends at that of the arrival census tract. These driving cost calculations contribute to evaluating the total expenses related to complete door-to-door driving trips and the driving segments of journeys that include multiple modes, such as trips to and from airports.

 \subsubsection{Ground Transportation}
 The standard mileage rate is a set amount per mile that taxpayers in the United States can use to calculate vehicle expenses for business, charitable, medical, or moving purposes instead of tracking actual costs like fuel, maintenance, and repairs. This rate simplifies record-keeping and expense reporting for individuals using their vehicles for these purposes. The costs of driving are calculated using the standard mileage rates provided by the United States Internal Revenue Services\footnote{Internal Revenue Services (IRS): \url{ https://www.irs.gov/tax-professionals/standard-mileage-rates }.Last accessed on [July 29\textsuperscript{th} 2024]}, coupled with driving distance estimates derived from the OSRM.

\subsubsection{AAM}
The cost of the ground transportation portion of the trip can be calculated from the above mentioned steps. But the costs tied to air travel hinge on various factors including distance, service class, booking lead time, time of day, day of the week, competitive presence, and airline concentration at the point of origin or destination. This study does not delve into modeling airfares across multiple explanatory variables but instead adopts a straightforward model that correlates passenger cost with distance. Data for this model is sourced from the DB1B database maintained by the Department Of Transportation (DOT), which contains a 10\% sample of all tickets sold in the United States for the years 2021 and 2022. This database comprises three components: the ticket, market, and coupon databases. For this analysis, the market database is employed to extract airfare data by filtering the flights based on their destination or arrival in the Tennessee state.

\subsection{Time Modeling}
Another crucial element in the decision-making process of travelers is the evaluation of the travel time linked with each mode of transportation. Accurately assessing the time savings in travel is essential to highlight the appeal of air transportation compared to ground-based options.
\subsubsection{Ground Transportation}
Driving times are calculated using the OSRM. For door-to-door journeys via automobile, the starting point and endpoint are assumed to be the population centroid of the origin census tract and the destination census tract, respectively.

\subsubsection{AAM}
The total travel time comprises several components: the drive from the origin to the departure airport, the duration spent at the departure airport, the flight time between the departure and destination airports (which may include layovers), the time spent at the destination airport, and finally, the drive to the final destination. The time for the first and last part of the trip which are ground transportation can be obtained from the above mentioned methods using the OSRM. For the aviation time we  perform the block time calculation for its modeling.

The block times are the periods from when the aircraft's brakes are released at departure to when they are set upon arrival, are regressed from the two years 2021 and 2022 from Aviation System Performance Metrics (ASPM) dataset\footnote{FAA Aviation System Performance Metrics (ASPM): \url{ https://aspm.faa.gov/apm/sys/AnalysisCP.asp }. Last accessed on [May 9\textsuperscript{th} 2024]}. The ASPM dataset is accessible through an online access system provided by the Federal Aviation Administration(FAA) and delivers comprehensive data regarding flights to and from the ASPM airports, as well as all flights operated by ASPM carriers. We used its data for city pair analysis, determining block time duration.

\subsection{Risk Modeling}
Risk modeling plays a crucial role in assessing and incorporating the various risks associated with different modes of transportation. The aim is to quantify the risk factors and include them in the overall cost calculation, thereby providing a more comprehensive estimate of the true cost of a trip. In our case we  calculate the number of fatalities for each mode of transportation.
The Value of Statistical Life (VSL) is an economic measure used to quantify the benefit of reducing the risk of death. It represents the monetary value society is willing to pay to save a life \cite{viscusi2003value}. With the help of those two we can find the monetary equivalent of the risk involved during each mode of transportation calculated as:
\begin{equation}
\label{equation:Risk}
R_m= VSL * \Lambda_m
\end{equation}
where
\begin{itemize}
    \item $m$ is mode of transportation either ground or AAM
    \item $R$  is risk of trip 
    \item $VSL$  is the amount society is willing to pay to save a life
    \item $\Lambda$ is number of fatalities per mile
\end{itemize}
Given that the dataset spans the years 2021-2022, the dataset for the VSL from those two years is obtained and averaged to determine the required VSL for this study from the US DOT\cite{USDOT2024}.

\subsubsection{Ground Transportation}
When analyzing motor-vehicle fatality trends across states, various fatality rate estimates can be employed to evaluate relative risk. Like calculating the total number of motor-vehicle deaths among state residents per 100,000 population, calculating the number of traffic deaths occurring in a state per 100 million vehicle miles traveled or calculating the number of traffic deaths per 10,000 registered vehicles. These three rate calculations offer distinct perspectives on fatality risk among states. In our study we used the second one which is based on the fatalities per 100 million vehicle miles travel. For the ground transportation parameter $\Lambda_G$ is obtained from the dataset \footnote{Motor-Vehicle Deaths by State:\url{ https://injuryfacts.nsc.org/state-data/motor-vehicle-deaths-by-state/ }. Last accessed on [May 9\textsuperscript{th} 2024]} provided by National Safety Council(NSC).

\subsubsection{AAM}
For the ground transportation section of the trip above mentioned methods can be deployed for the risk calculation. For commercial scheduled air travel, which is among the safest modes of transportation, the lifetime odds of a fatality for an aircraft passenger in the United States are negligible. For our study to calculate the number of fatalities for air transportation per million miles from the airplane crashes dataset\footnote{Airplane Crashes:\url{ https://injuryfacts.nsc.org/home-and-community/safety-topics/airplane-crashes/ }. Last accessed on [May 9\textsuperscript{th} 2024]} provided by NSC.

\subsection{Generalized Cost Modeling}
To model the mode choice of travelers, researchers often begin with disaggregate models that estimate the utility of each transportation mode, factoring in attributes of both the travelers and the transportation modes. However, calibrating utility functions with numerous attributes can be complex and often requires extensive surveys. To simplify this, study \cite{roy2021user} proposed using the concept of GCT, which considers two main components: the monetary cost ($C_m$) of the trip and the opportunity cost, calculated as the traveler's value of time ($W$) multiplied by the door-to-door journey time ($T_m$). The monetary cost includes direct expenditures such as fuel or ticket prices, while the opportunity cost reflects the economic value of time spent traveling calculated by averaging the median hourly wages\footnote{U.S. BUREAU OF LABOR STATISTICS: \url{ https://www.bls.gov/oes/tables.htm }. Last accessed on [May 9\textsuperscript{th} 2024]} in the census tracts of origin and destination. The negative sign preceding the individual terms in the equation (3), corresponds to the disutility associated with the increase in trip cost, trip time and risk involved during the trip. 
\begin{equation}
\label{equation:GCT}
GCT_m= -C_m - W*T_m - R_m
\end{equation}
where
\begin{itemize}
    \item $m$ is mode of transportation either ground or AAM
    \item $GCT$  is  generalized cost of trip
    \item $C$  is cost of trip per mile per passenger
    \item $W$ is average median hourly wage of origin and destination census tracts
    \item $T$ is time of trip 
    \item $R$ is risk of trip 
\end{itemize}

\subsection{Mode Choice}
The GCT metric is suitable for a deterministic choice model, which assumes that individuals are perfectly rational in their decision-making process. However, this approach may overlook other aspects of individual decision-making that analysts do not fully understand. To account for these uncertainties, a probabilistic-based choice model is more desirable, where the probability of choosing a transportation mode is considered instead of a deterministic choice. This probabilistic model represents the true utility as the sum of the systematic component of the utility and a random error term.
\begin{equation}
\label{equation:Utility}
\begin{aligned}
U_{\text{G}} &=  GCT_{\text{G}} + \epsilon; \\
U_{\text{AAM}} &=  GCT_{\text{AAM}} + \epsilon
\end{aligned} 
\end{equation}
where
\begin{itemize}
    \item $U$ is utility for trip
    \item $GCT$  is  generalized cost of trip
    \item $\epsilon$  is the error associated
\end{itemize}
The behavior of travelers when faced with transportation mode options is simulated using discrete choice models, which estimate the probability of selecting a specific mode. These models, such as the logit and conditional logit models, vary based on assumptions about the distribution of the error component ($\epsilon$) of the utility model. A common assumption is that the error term follows an independent and identically distributed (IID) Gumbel distribution, resulting in the multinomial logit model\cite{train2009discrete}. This model provides a closed-form expression for the choice probability as a function of the differences between the utilities of the transportation modes. By incorporating the variability in individual preferences through the error term, discrete choice models offer a probabilistic framework that more accurately reflects real-world decision-making. The probability of choosing transportation mode(air or ground) can be expressed as: 
\begin{equation}
\label{equation:Probability}
\begin{aligned}
P_{\text{AAM}} = \frac{1}{1 + e^{(U_{\text{G}} - U_{\text{AAM}})}} \\
P_{\text{AAM}} = \frac{1}{1 + e^{ (GCT_{\text{G}} - GCT_{\text{AAM}})}}
\end{aligned}
\end{equation}
where
\begin{itemize}
    \item $P_{\text{AAM}}$ is Probability of selecting AAM
    \item $U$ is utility for trip
    \item $GCT$  is  generalized cost of trip
\end{itemize}

\section{Evaluation and Discussion}
\subsection{Cost Modeling}
The regression analysis depicted in \textcolor{blue}{Figure}\autoref{fig:Cost} illustrates the relationship between passenger cost per mile and travel distance for air travel within Tennessee. The analysis reveals a clear inverse relationship between cost per mile and travel distance, indicating that shorter flights tend to have a higher cost per mile compared to longer flights. This trend aligns with the economies of scale in air travel, where fixed costs are distributed over a greater number of miles in longer flights, thereby reducing the cost per mile. Also, the variability in costs is more pronounced for shorter distances, likely due to factors such as demand fluctuations, service class variations, and competition.


\subsection{Time Modeling}
Research on block time in various industries has shown that it tends to increase with distance, but not exponentially\cite{fan2019schedule}. In the airline industry, scheduled block times have been found to grow more linearly or with some polynomial relationship, influenced by factors such as air traffic growth, airport congestion, flight delays,  takeoff, landing, and cruising times. Keeping this context under consideration we employ polynomial fit for the time modeling.

The regression analysis illustrated in the \textcolor{blue}{Figure}\autoref{fig:BlockTime} examines the relationship between block time (in hours) and travel distance (in miles) for air travel. The scatterplot, showing actual block times (blue dots) and the regressed block time (red line), reveals a positive correlation between block time and distance, indicating that longer flights generally require more time. However, there is a significant variability in block times, especially for shorter distances, likely due to factors such as airport congestion, layovers, and flight scheduling irregularities. The regression line shows that block time increases with distance but not linearly, reflecting efficiency gains on longer flights. Notably, around distances of 480 miles, block time increases more rapidly, possibly due to the inclusion of more itineraries with layovers, adding extra time at transit airports. While block time generally rises with travel distance, operational factors contribute to considerable variability.


\subsection{Risk Modeling}
\begin{figure}[hbt!]
\centering
\includegraphics[width=\columnwidth]{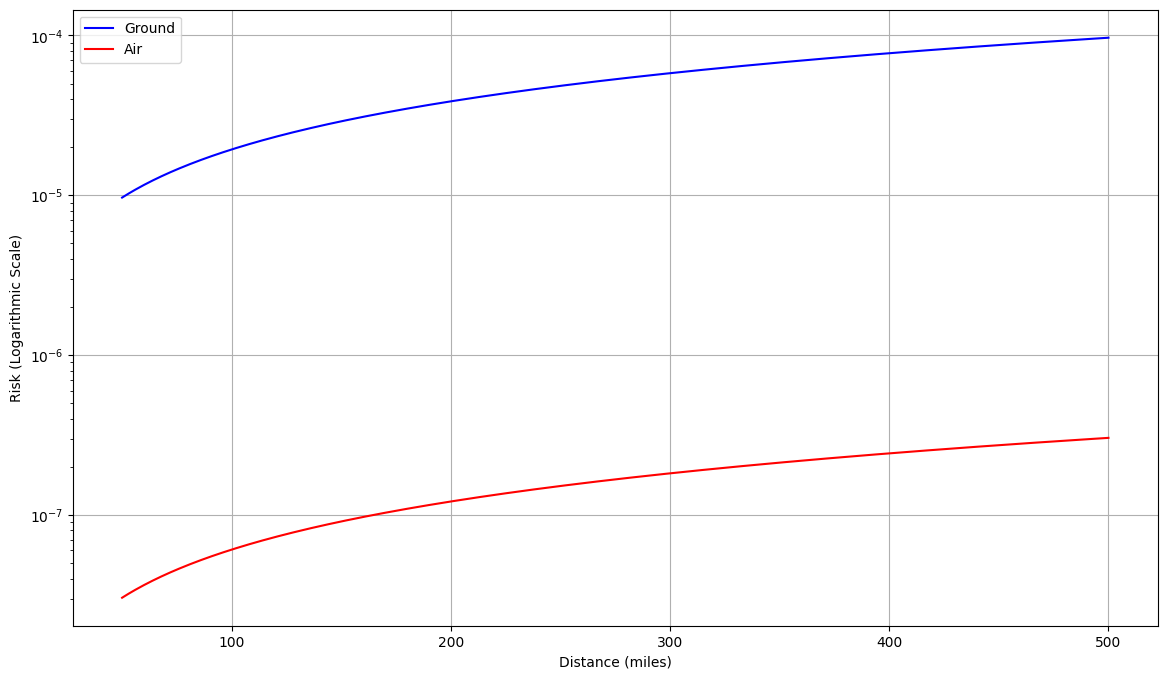}
\caption{Risk modeling for different modes of Transportation}
\label{fig:risk}
\end{figure}

The analysis shown in \autoref{fig:risk} indicates that ground transportation presents a significantly higher fatality risk compared to air travel over distances up to 500 miles, with risks increasing more sharply for ground travel. The logarithmic scale of the graph illustrates that ground travel risk starts around \(10^{-5}\) and approaches \(10^{-4}\), while air travel risk remains substantially lower, starting around \(10^{-7}\) and barely exceeding \(10^{-6}\). This considerable disparity underscores the relative safety of air travel, even though both modes show increasing risks with distance. However, the monetary values of the risk associated with both transportation modes are negligible compared to other costs incurred, such as time and actual expenses.

\subsection{Generalized Cost Modeling}

\begin{figure}[hbt!]
\centering
\includegraphics[width=\columnwidth]{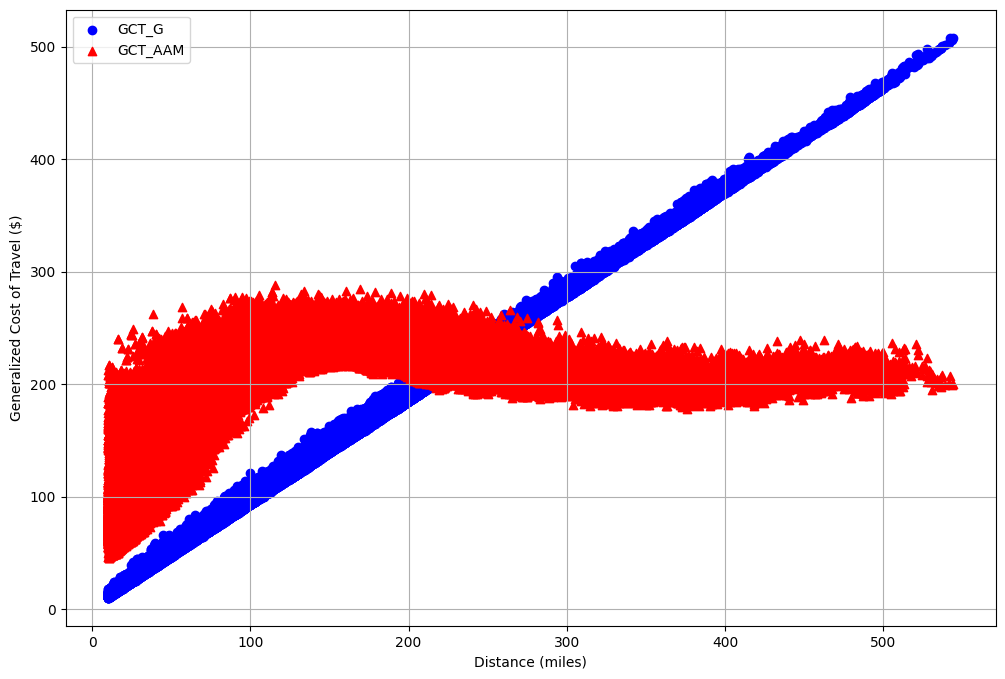}
\caption{Regression Analysis For Generalized Cost of Trip}
\label{fig:GCT}
\end{figure}

The regression analysis depicted in \autoref{fig:GCT} examines the relationship between the GCT in dollars and travel distance in miles for two transportation modes: AAM and ground transportation. The analysis shows that the GCT for ground transportation increases linearly with distance, indicating a consistent rise in costs as travel distance extends. In contrast, the GCT for AAM exhibits a non-linear pattern, initially increasing, then slightly decreasing, and finally stabilizing, suggesting that AAM travel may become more cost-efficient over certain distances. This pattern likely reflects the fixed costs and economies of scale associated with AAM travel, which become more pronounced over longer distances. For longer distances, the GCT for AAM stabalizes around 200.

\subsection{Mode Choice}
\begin{figure}[hbt!]
\centering
\includegraphics[width=\columnwidth]{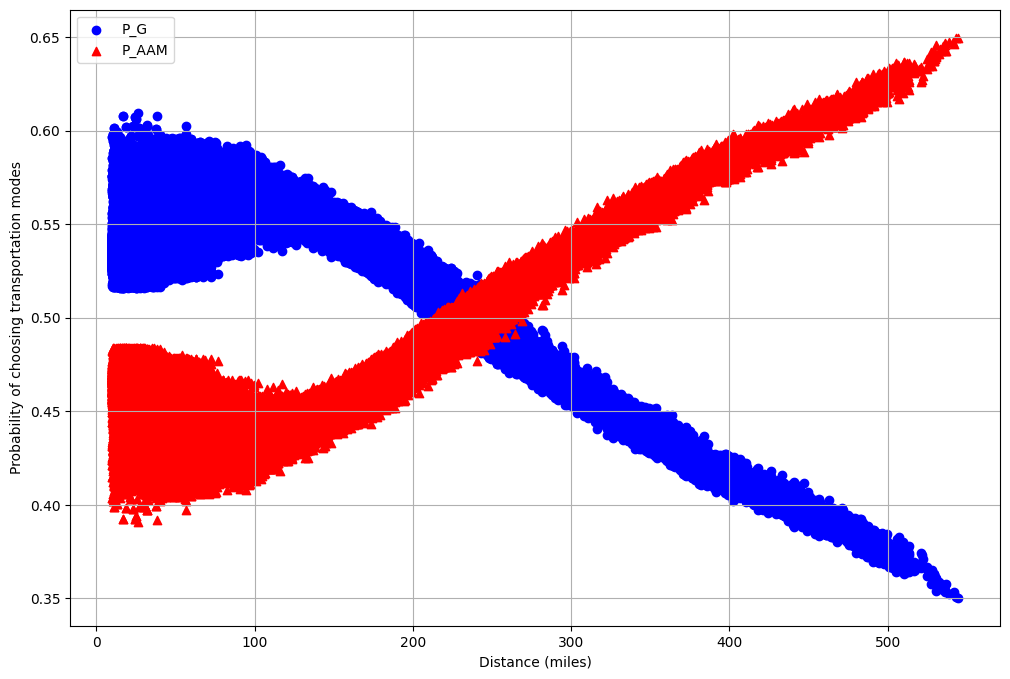}
\caption{Probability of choosing transportation mode}
\label{fig:Probability}
\end{figure}

\begin{figure}[hbt!]
\centering
\includegraphics[width=\columnwidth]{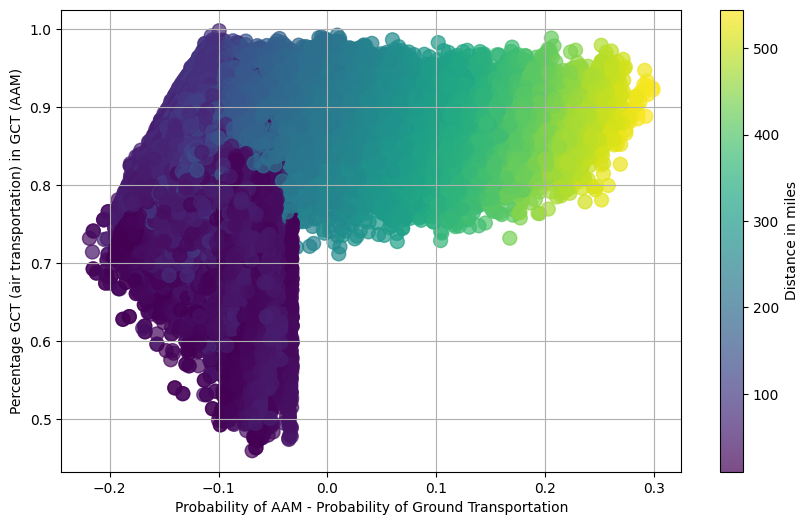}
\caption{Percentage of $GCT_A$ in $GCT_{AAM}$ vs difference of $P_G$ and $P_{AAM}$}
\label{fig:modechoicePercentage}
\end{figure}

We performed the calculation for finding the probability of choosing the mode of transportation between AAM and ground transportation using equation (5).  The \autoref{fig:Probability} depicts the likelihood of selecting various transportation modes relative to distance. Initially, as distance increases, the probability of opting for ground transportation remains higher but then progressively declines, indicating a preference for AAM over longer distances. For distances up to approximately 200 miles, ground transportation is more frequently chosen. However, beyond around 250 miles, AAM becomes the favored mode. This pattern reveals a critical distance threshold where the likelihood of choosing AAM exceeds that of ground transportation, emphasizing its increasing appeal for extended trips. The \autoref{fig:modechoicePercentage} shows that apart from the threshold distance, the demand for AAM is also higher in trips where the percentage of GCT of air transportation exceeds 70\% of the total GCT for AAM.

\begin{table*}[h]
\centering
\caption{\label{tab:results} Mean Values for AAM and Non-AAM Trips}
\begin{tabular}{|c|c|c|}
\hline
\textbf{Variable}         & \textbf{Non-AAM} & \textbf{AAM} \\ \hline
GCT by Air Transportation (\$)              & 151.07                  & 182.16              \\ \hline
GCT by Ground Transportation   (\$)            & 27.30                   & 24.96               \\ \hline
Time in Ground Transportation (hours)             & 0.73                    & 0.67                \\ \hline
Distance by Ground Transportation (miles)         & 23.39                   & 21.40               \\ \hline
Distance by Air Transportation   (miles)           & 72.98                   & 273.19              \\ \hline
Time in Air Transportation   (hours)         & 0.58                    & 1.18                \\ \hline
Distance between OD (miles)         & 87.88                   & 318.53              \\ \hline
Ground Transportation time between OD    (hours)           & 1.86                    & 6.06                \\ \hline
\end{tabular}
\end{table*}

The \autoref{tab:results} presents comparative data on the average characteristics of trips involving AAM and non-AAM modes of transportation. It is evident that AAM trips, on average, exhibit larger GCT in air transportation (\$182.16 ) less in GCT by ground transportation (\$24.96) compared to non-AAM trips which is \$151.07 and \$27.30 in air and ground transportation respectively, highlighting the extended air travel associated with AAM. Ground transportation distances and time durations are reduced in AAM trips (21.40 miles and 0.67 hours) compared to Non-AAM trips (0.73 hours and 23.39 miles), suggesting less ground segment connectivity in AAM networks. However, AAM trips demonstrate significantly longer distances  and time duration by air (273.19 miles and 1.18 hours) than Non-AAM trips (72.98 miles and 0.58 hours), indicating that AAM is potentially more suited for longer-distance travel where traditional ground travel is deemed inefficient. Furthermore, average distance between origins and destinations is 318.53 miles in comparison to 87.88 miles for non-AAM trips.

\autoref{tab:Analysis}  presents a comparison of trip demand features between all trips and those attributed to AAM. The data is broken down by age, earnings, and industry sectors. For age, individuals aged 30-54 make up the largest percentage of both, indicating a dominant demographic in travel demand. In terms of earnings, those making over \$3333/month represent the largest share for both trip categories, suggesting higher-income individuals are frequent travelers. Industry-wise, the 'Other Services industry' holds the highest percentage for both trip types.

\renewcommand{\arraystretch}{1.15} 
\begin{table*}[htbp]
\caption{\label{tab:Analysis} Comparison of Trip Demand Features}
\begin{center}
\begin{tabular}{|c|l|l|l|}
\hline
\multicolumn{1}{|c|}{\textbf{Features}} & \multicolumn{1}{c|}{\textbf{Sub-categories}} & \multicolumn{1}{c|}{\textbf{All Trips (\%)}} & \multicolumn{1}{c|}{\textbf{AAM Trips (\%)}} \\ \hline

\multirow{3}{*}{\textbf{Age}} 
& 29 or younger & 23.47 & 22.88 \\ \cline{2-4}
& 30 - 54 & 53.28 & 53.34 \\ \cline{2-4}
& 55 or Older & 23.25 & 23.78 \\ \hline

\multirow{3}{*}{\textbf{Earning}} 
& \$1250/month or less & 23.51 & 26.59 \\ \cline{2-4}
& \$1251/month to \$3333/month & 32.06 & 29.49 \\ \cline{2-4}
& Greater than \$3333/month & 44.43 & 43.92 \\ \hline

\multirow{3}{*}{\textbf{Industry}} 
& Goods Producing & 16.91 & 10.89 \\ \cline{2-4}
& Trade, Transportation, and Utilities & 21.61 & 33.75 \\ \cline{2-4}
& Other Services industry & 61.48 & 55.36 \\ \hline

\end{tabular}
\end{center}
\end{table*}


 
\begin{figure*}[hbt!]
    \centering
    \subfloat[Job part of OD pairs\label{fig:jobsCensus}]{
        \includegraphics[width=0.98\columnwidth]{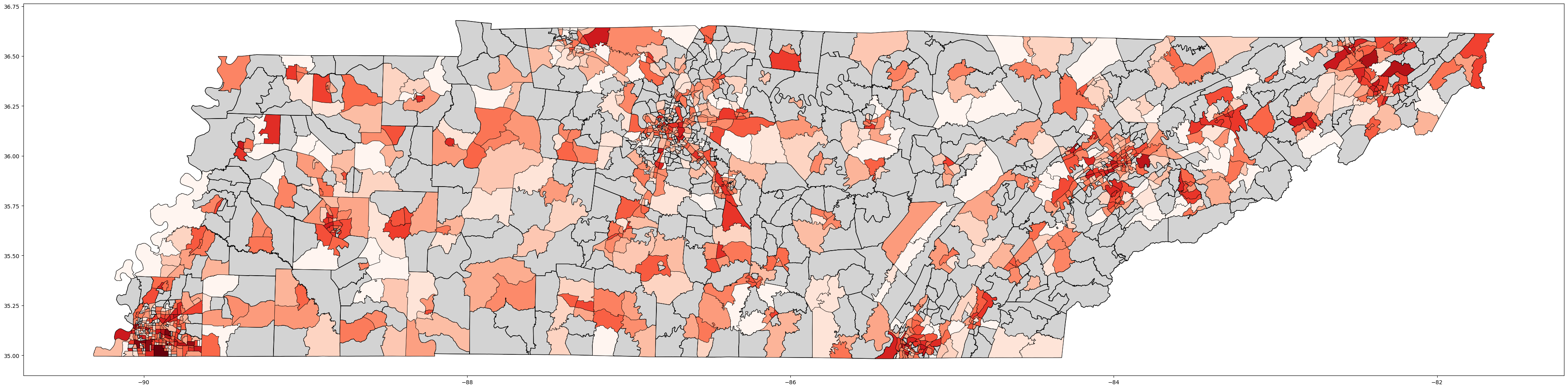}
    }
    \hfill
    \subfloat[Home part of OD pairs\label{fig:homeCensus}]{
        \includegraphics[width=0.98\columnwidth]{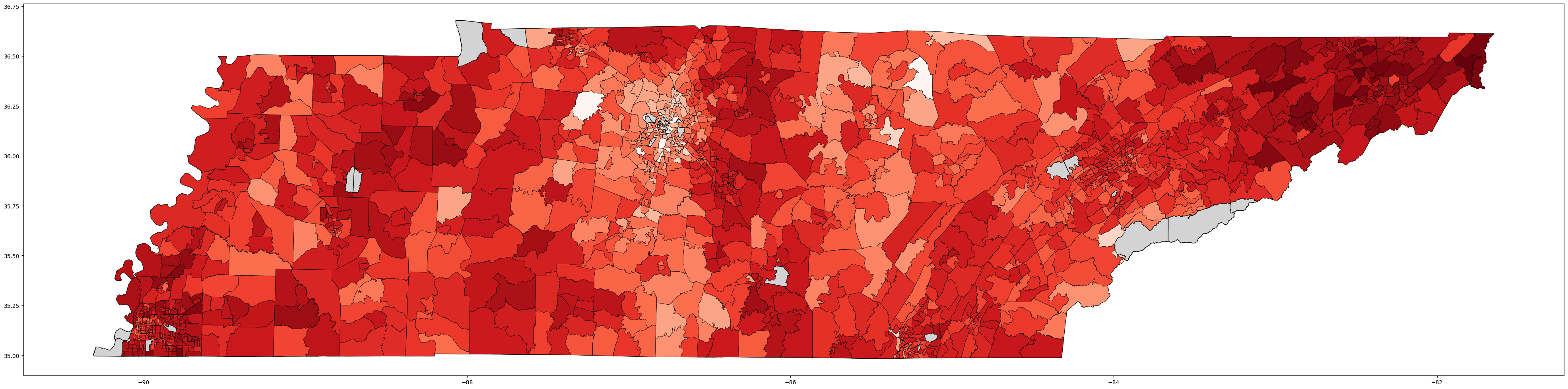}
    }
    \caption{Census Tracts for AAM OD pairs in Tennessee}
    \label{fig:ODPairs}
\end{figure*}

The \autoref{fig:ODPairs} illustrate distinct distributions for job and home locations within the Tennessee state's census tracts. Jobs are densely concentrated in central Tennessee, particularly around Nashville, as well as in the eastern region near Knoxville, and in the southwestern area near Memphis, indicating major urban centers with high employment opportunities. In contrast, residential areas, while also concentrated around these cities, display a broader spread across the region, suggesting a more dispersed living pattern typical of suburban and extended urban areas. This distribution pattern highlights the centralization of employment in urban hubs and the wider spread of residences, which likely influences commuting flows in Tennessee.

\section{Conclusion}

In this research paper, the cost, time, and risk models were developed for AAM and ground transportation. These models were subsequently utilized to create the GCT model to predict the likelihood of AAM being chosen as the preferred mode of transportation. The proposed model indicated that, for Tennessee, trip demand generated at the census tract level is more likely to constitute AAM trip demand if air transportation contributes more than 70\% to the GCT and if the trip distance exceeds 250 miles. Future research direction will be to estimate the costs for AAM aircraft powered by electric charge and its impact on the GCT and selection of AAM as preferred mode.

\section*{Acknowledgment}
This material is based upon work supported by the NASA Aeronautics Research Mission Directorate (ARMD) University Leadership Initiative (ULI) under cooperative agreement number 80NSSC23M0059. This research was also partially supported by the U.S. National Science Foundation through Grant No. 2317117 and Grant No. 2309760.

\bibliographystyle{IEEEtran}
\bibliography{ref.bib}

\begin{thebibliography}{10}
\providecommand{\url}[1]{#1}
\csname url@samestyle\endcsname
\providecommand{\newblock}{\relax}
\providecommand{\bibinfo}[2]{#2}
\providecommand{\BIBentrySTDinterwordspacing}{\spaceskip=0pt\relax}
\providecommand{\BIBentryALTinterwordstretchfactor}{4}
\providecommand{\BIBentryALTinterwordspacing}{\spaceskip=\fontdimen2\font plus
\BIBentryALTinterwordstretchfactor\fontdimen3\font minus \fontdimen4\font\relax}
\providecommand{\BIBforeignlanguage}[2]{{%
\expandafter\ifx\csname l@#1\endcsname\relax
\typeout{** WARNING: IEEEtran.bst: No hyphenation pattern has been}%
\typeout{** loaded for the language `#1'. Using the pattern for}%
\typeout{** the default language instead.}%
\else
\language=\csname l@#1\endcsname
\fi
#2}}
\providecommand{\BIBdecl}{\relax}
\BIBdecl

\bibitem{antcliff2021regional}
K.~Antcliff, N.~Borer, S.~Sartorius, P.~Saleh, R.~Rose, M.~Gariel, J.~Oldham, C.~Courtin, M.~Bradley, S.~Roy \emph{et~al.}, ``Regional air mobility: Leveraging our national investments to energize the american travel experience,'' 2021.

\bibitem{mckinseyShorthaulFlying}
L.~Brink, R.~Brown, S.~Carter, A.~Esqué, B.~Meigs, and R.~Riedel, ``{S}hort-haul flying redefined: {T}he promise of regional air mobility --- mckinsey.com,'' \url{https://www.mckinsey.com/industries/aerospace-and-defense/our-insights/short-haul-flying-redefined-the-promise-of-regional-air-mobility}, 2023, [Accessed 10-06-2024].

\bibitem{rajendran2019insights}
S.~Rajendran and J.~Zack, ``Insights on strategic air taxi network infrastructure locations using an iterative constrained clustering approach,'' \emph{Transportation Research Part E: Logistics and Transportation Review}, vol. 128, pp. 470--505, 2019.

\bibitem{holden2016fast}
J.~Holden and N.~Goel, ``Fast-forwarding to a future of on-demand urban air transportation,'' \emph{San Francisco, CA}, 2016.

\bibitem{davies1979cluster}
D.~L. Davies and D.~W. Bouldin, ``A cluster separation measure,'' \emph{IEEE transactions on pattern analysis and machine intelligence}, no.~2, pp. 224--227, 1979.

\bibitem{rajendran2021predicting}
S.~Rajendran, S.~Srinivas, and T.~Grimshaw, ``Predicting demand for air taxi urban aviation services using machine learning algorithms,'' \emph{Journal of Air Transport Management}, vol.~92, p. 102043, 2021.

\bibitem{ahmed2024demand}
F.~Ahmed, M.~A. Memon, K.~Rajab, H.~Alshahrani, M.~E. Abdalla, A.~Rajab, R.~Houe, and A.~Shaikh, ``Demand prediction for urban air mobility using deep learning,'' \emph{PeerJ Computer Science}, vol.~10, p. e1946, 2024.

\bibitem{tarafdar2019urban}
S.~Tarafdar, M.~Rimjha, N.~Hinze, S.~Hotle, and A.~A. Trani, ``Urban air mobility regional landing site feasibility and fare model analysis in the greater northern california region,'' in \emph{2019 Integrated Communications, Navigation and Surveillance Conference (ICNS)}.\hskip 1em plus 0.5em minus 0.4em\relax IEEE, 2019, pp. 1--11.

\bibitem{mcnally2007four}
M.~McNally, ``The four-step model. chapter 3 in hensher d. and button k.(eds). handbook of transport modeling”,'' 2007.

\bibitem{robusto1957cosine}
C.~C. Robusto, ``The cosine-haversine formula,'' \emph{The American Mathematical Monthly}, vol.~64, no.~1, pp. 38--40, 1957.

\bibitem{viscusi2003value}
W.~K. Viscusi and J.~E. Aldy, ``The value of a statistical life: a critical review of market estimates throughout the world,'' \emph{Journal of risk and uncertainty}, vol.~27, pp. 5--76, 2003.

\bibitem{USDOT2024}
{U.S. Department of Transportation}, ``Revised departmental guidance on valuation of a statistical life in economic analysis,'' \url{https://www.transportation.gov/office-policy/transportation-policy/revised-departmental-guidance-on-valuation-of-a-statistical-life-in-economic-analysis}, 2024, accessed: 2024-06-10.

\bibitem{roy2021user}
S.~Roy, M.~T. Kotwicz~Herniczek, B.~J. German, and L.~A. Garrow, ``User base estimation methodology for a business airport shuttle air taxi service,'' \emph{Journal of Air Transportation}, vol.~29, no.~2, pp. 69--79, 2021.

\bibitem{train2009discrete}
K.~E. Train, \emph{Discrete choice methods with simulation}.\hskip 1em plus 0.5em minus 0.4em\relax Cambridge university press, 2009.

\bibitem{fan2019schedule}
T.~P.~C. Fan, ``Schedule creep--in search of an uncongested baseline block time by examining scheduled flight block times worldwide 1986--2016,'' \emph{Transportation Research Part A: Policy and Practice}, vol. 121, pp. 192--217, 2019.

\end{thebibliography}

\end{document}